\documentclass[twocolumn,prl,amssymb,superscriptaddress]{revtex4-1}

\usepackage{graphicx}
\usepackage{amsmath}
\usepackage{amsfonts}
\usepackage{amsthm}
\usepackage{amssymb}
\usepackage{amsbsy}
\usepackage{wasysym}
\usepackage{bbm}
\usepackage{bm}
\usepackage{mathrsfs}
\usepackage{color}
\usepackage{hyperref}
\usepackage{braket}

\setcitestyle{super}

\newcommand*{\citen}[1]{%
  \begingroup
    \romannumeral-`\x 
    \setcitestyle{numbers}%
    \cite{#1}%
  \endgroup   
}

\date{\today}
\begin{document}
\title{Kardar-Parisi-Zhang universality from soft gauge modes}
\author{Vir B. Bulchandani}
\affiliation{Department of Physics, University of California, Berkeley, Berkeley CA 94720, USA}

\begin{abstract}
The emergence of superdiffusive spin dynamics in integrable classical and quantum magnets is well established by now, but there is no generally valid theoretical explanation for this phenomenon. A fundamental difficulty is that the hydrodynamic fluctuations of conserved quasiparticle modes are purely diffusive. We argue that in isotropic integrable magnets, a complete hydrodynamic description must include soft ``gauge'' degrees of freedom, that arise from spontaneous breaking of the Bethe pseudovacuum symmetry. We show that the coarse-grained time evolution of these modes lies in the Kardar-Parisi-Zhang universality class of dynamics.
\end{abstract}

\maketitle

\paragraph{Introduction.} The evolution of chaotic, many-particle systems on long length and time scales is described by hydrodynamics, which gives rise to universal behaviours that are largely independent of microscopic details and determined by a small number of local conservation laws. In low dimensions, the classification of such universal behaviours is complicated by the possibility of anomalous transport, characterized by divergent linear-response coefficients\cite{AW}. These difficulties are most pronounced in one dimension, where perturbative techniques break down entirely. Despite such obstacles, the past decade has seen substantial progress in the hydrodynamics of classical, chaotic, one-dimensional systems, through the application of techniques from stochastic field theory within the framework of nonlinear fluctuating hydrodynamics\cite{vanB,SpohnFluct,KulkLam,Fib}.

Another unusual feature of hydrodynamics in one dimension is the existence of integrable many-body systems, such as the classical Toda anharmonic chain and the quantum spin-$1/2$ Heisenberg magnet. Such systems possess extensive families of local conservation laws, that prevent them from thermalizing in the conventional sense\cite{Rigol,BS} and place them beyond standard hydrodynamic techniques. Although the kinetic theory of certain classically integrable many-body systems, such as soliton gases, has been known for some time\cite{Percus,Zakharov,Boldrighini,EK}, a unified theory of the hydrodynamics of integrable systems, both quantum and classical, was only achieved recently \cite{Bertini2016,Castro-Alvaredo2016,Bulchandani2018}. The resulting generalized hydrodynamics of integrable systems has been thoroughly verified through numerical simulation of a wide range of quantum and classical systems, and in a cold-atom realization of the Lieb-Liniger gas (see e.g. \citen{Schemmer} and references therein).

One numerically observed phenomenon whose explanation has remained beyond both of these theoretical advances is the emergence of Kardar-Parisi-Zhang (KPZ) universality in the dynamics of classical and quantum integrable systems with isotropic rotational symmetry\cite{KPZ1,KPZ2,KPZ3,KPZ4,KPZ5,KPZ5b,KPZ6,KPZ7}. The essential theoretical difficulty is that when the methods of nonlinear fluctuating hydrodynamics are applied to the evolution of quasiparticle modes, as described by generalized hydrodynamics, they predict purely diffusive fluctuations\cite{KPZ5,DoyonFluct}. While for the spin-$1/2$ Heisenberg model, it is possible to explain the emergence of a $z=3/2$ dynamical exponent self-consistently within generalized hydrodynamics\cite{GV}, and semiclassical arguments have been put forward for KPZ scaling in low-temperature quantum antiferromagnets\cite{KPZ5}, neither of these explanations has sufficient scope to address the apparently universal origin of KPZ scaling in isotropic integrable magnets\cite{KPZ6,KPZ7}.

This paper points out a generic mechanism for the emergence of such physics; the key observation is that for isotropic, integrable systems, the generalized hydrodynamics of infinitely many conservation laws is not equivalent to the Bethe-Boltzmann equation describing the propagation of quasiparticle modes. The long-wavelength dynamics of the two spin degrees of freedom transverse to the local magnetization can decouple from the short-wavelength quasiparticle modes, and it is the fluctuating hydrodynamics of these soft, transverse degrees of freedom that gives rise to the observed Kardar-Parisi-Zhang scaling.

The paper is laid out as follows. We first derive the result that fluctuations of quasiparticle mode occupancies scale diffusively. This fact appears to be known\cite{KPZ5,DoyonFluct}, although its derivation has not been published, and since it is essential for our argument we derive it below. We next discuss the treatment of magnetization in thermodynamic Bethe ansatz, and show that at the level of hydrodynamics, it allows for long-wavelength gauge modes that evolve according to the Landau-Lifshitz equation. We then demonstrate that the coarse-grained Landau-Lifshitz dynamics yields a conserved torsional degree of freedom, that evolves in time as the derivative of a Kardar-Parisi-Zhang height function, providing the first generic mechanism that explains the numerical findings in previous works. We also propose some signatures of this torsional degree of freedom that could be tested in quantum or classical numerical simulations. In the final discussion, we explain why the emergence of KPZ scaling requires both integrability and isotropic symmetry to be robust at long times, as indicated by recent numerical results\cite{KPZ6,KPZ7}.
 
\paragraph{Nonlinear fluctuating hydrodynamics of quasiparticle modes.} We first show that within the generalized hydrodynamics of integrable systems, fluctuations of quasiparticle modes with pseudomomentum $k$ all have a diffusive dynamical exponent, $z_k =2$, following the method of nonlinear fluctuating hydrodynamics reviewed in Ref. \citen{SpohnFluct}. Our starting point is the Bethe-Boltzmann equation for the evolution of quasiparticle mode occupancies in integrable systems\cite{Bertini2016,Castro-Alvaredo2016},
\begin{equation}
\label{bb}
\partial_t \rho_k + \partial_x (\rho_k v_k[\rho]) = 0.
\end{equation}
Here, $\rho_k$ is the density of occupied states, $v_k[\rho]$ is the quasiparticle velocity, and we shall take $\rho^t_k$ and $\theta_k$ to denote the total density of states and Fermi factor respectively (We suppress a possible quasiparticle flavour index for economy of notation.) Linearizing about an equilibrium configuration $\rho^{eq}_k$ yields
\begin{equation}
\partial_t \delta \rho_k + \partial_x (\hat{A}\delta \hat{\rho})_k = 0,
\end{equation}
where the linearized velocity operator is
\begin{equation}
\hat{A} =  (\hat{1}+\hat{\theta}\hat{K})^{-1}\hat{v}(1+\hat{\theta}\hat{K}).
\end{equation}
(Here $K$ denotes the differential phase shift and $\hat{O}$ denotes the operator corresponding to the integral kernel $O_{kk'}$.) We next consider the equilibrium correlation matrix $C_{kk'} = \langle \delta \rho_k \delta \rho_{k'} \rangle_{\rho^{eq}}$. By thermodynamic stability, this vanishes for an infinite system. The natural length scale to choose here is the hydrodynamic coarse-graining length $l$, which yields\cite{FendleySaleur}
\begin{equation}
C_{kk'} = \frac{1}{l} \left[(1+\hat{\theta}\hat{K})^{-1}\hat{\rho}(1-\hat{\theta})(1+\hat{K}\hat{\theta})^{-1} \right]_{kk'}.
\end{equation}
The sum rule $\hat{A}\hat{C} = \hat{C}\hat{A}^T$ is immediate\cite{DoyonFluct}. Introducing noise and an associated dissipative term, we have
\begin{equation}
\label{rholin}
\partial_t \delta \rho_k + \partial_x \left[(\hat{A} \delta \hat{\rho})_k -(\hat{D}\partial_x \delta \hat{\rho})_k + (\hat{B}\hat{\zeta})_{k}\right] = 0,
\end{equation}
with $\zeta_k$ a white noise with zero mean and unit variance,  $\quad \mathbb{E}[\zeta_k(x,t) \zeta_{k'}(x',t')] = \delta(x-x')\delta(k-k') \delta(t-t')$. Stationarity of correlations $\mathbb{E}[\delta \rho_k(x,t) \delta \rho_{k'}(0,0)]$ with respect to the evolution Eq. \eqref{rholin} imposes the fluctuation-dissipation relation $\hat{D}\hat{C} + \hat{C}\hat{D}^T =  \hat{B}\hat{B}^T$. We note that there is no general way to specify $\hat{D}$ and $\hat{B}$ independently\cite{SpohnFluct}. One might ask how to interpret this ambiguity, in view of recent analytical predictions for the diffusion kernel $\hat{D}$ from generalized hydrodynamics\cite{Diff1,Diff2,Diff3}. On this point, we remark that the mathematical derivation of the diffusion kernel in the classical hard rod gas assumes instantaneous molecular chaos. Whether or not this assumption holds in practice depends on the microscopic details of the system in question, in particular the presence of sufficient noise below the hydrodynamic length scale. This was demonstrated explicitly in the first numerical test of the diffusion kernel formula for the classical hard rod gas\cite{Boldrighini2,DS}, where it was found that the diffusive term has more or less predictive power depending on the strength of the microscopic thermalization mechanism\cite{Cao2018}. 

Thus we should expect that $\hat{B} \neq 0$ for realistic systems and by the fluctuation-dissipation relation, the recent analytical results correspond to an infinite-size limit, in which the hydrodynamic coarse-graining length $l \to \infty$. This observation is important for the present discussion because restoring a finite hydrodynamic coarse-graining length regulates the divergence of the spin diffusion constant in the isotropic spin-$1/2$ Heisenberg chain\cite{DivD}; strings of length $n > l$ are \textit{a priori} excluded from the description of local equilibrium states (essentially the same observation can be used to explain the emergence of a $z=3/2$ dynamical exponent\cite{GV}.) In particular, this implies that for the isotropic chain, the component of spin transport captured by the Bethe-Boltzmann equation can be regarded as having a finite diffusion constant at the level of fluctuating hydrodynamics, and does not need to be treated separately.

We now address the quadratic corrections to Eq. \eqref{rholin}, which by renormalization group arguments, determine the dynamical exponent of fluctuations of quasiparticle modes\cite{SpohnFluct,Fib}. For brevity, we first change variables to normal modes, $\phi_k = \int dk' \, R_{kk'} \delta \rho_{k'}$. The normalization convention $\hat{R} \hat{C} \hat{R}^T = \hat{1}$ fixes
\begin{equation}
R_{kk'} = \sqrt{\frac{l}{\rho_k(1-\theta_k)}} (\hat{1}+ \hat{\theta} \hat{K})_{kk'}.
\end{equation}
Letting $\hat{D}' = \hat{R}\hat{D}\hat{R}^{-1}, \hat{B}' = \hat{R}\hat{B}$ and expanding the quasiparticle velocity, the leading nonlinear evolution is found to be
\begin{equation}
\label{normal}
\partial_t \phi_k + \partial_x\left[v_k \phi_k - (\hat{D}'\hat{\phi})_k + (\hat{B}'\hat{\zeta})_k+ (\hat{\phi}^T \hat{G}^k \hat{\phi}) \right] = 0,
\end{equation}
where in terms of the dressed kernel $\alpha = -(\hat{1}+\hat{K} \hat{\theta})^{-1}\hat{K}$ and writing $R_{kk'} = \lambda_kM_{kk'}$ with $\hat{M} = (\hat{1}+\hat{\theta}\hat{K})$, we have\footnote{Using the operator identity $\hat{M}^{-1} - \hat{\theta} \hat{\alpha} = \hat{1}$, one can write this in the simplified form $G^k_{k'k''} = - \frac{1}{2} \frac{\lambda_k}{\rho^t_k} \frac{\alpha_{k'k''}}{\lambda_{k'}\lambda_{k''}}(v_{k'}-v_{k''})(\delta_{kk'} - \delta_{kk''})$, which matches a subsequent calculation\cite{Subs} of $G$, up to a choice of normalization in the definition of $C$.}
\begin{align}
\nonumber G^{k}_{k'k''} = & \frac{1}{2}\frac{\lambda_k M_{kl}}{\lambda_{k'}\lambda_{k''}} \Big[\frac{\alpha_{lk'}}{\rho_l^t}(v_{k'}-v_{l})M^{-1}_{lk''} + (k' \leftrightarrow k'') \\
\nonumber + \theta_l &\Big(\frac{\alpha_{lk''}\alpha_{k'k''}}{\rho^t_{k''}}(v_{k'} - v_{k''}) + (k' \leftrightarrow k'') \\
& + \frac{\alpha_{lk'}\alpha_{lk''}}{\rho^t_l}(2v_l - v_{k'}-v_{k''}) \Big) \Big].
\end{align}
The dynamical universality class of fluctuations in Eq. \eqref{normal} is dictated by the diagonal elements $G^k_{k'k'}$ of the Hessian. It is easily verified that $G^k_{k'k'} = 0$, implying\cite{Fib} a dynamical exponent $z_k = 2$ for fluctuations about every quasiparticle mode $\phi_k$. Since the structure of the velocity derivatives leading to this result is essentially fixed by the semi-Hamiltonian geometry of Eq. \eqref{bb} (for example, the Christoffel symbols determine the dressed kernel), we expect purely diffusive fluctuations to be a generic property of integrable kinetic equations of this type\cite{Int1,Int2}. 

We emphasize that our derivation is ``mesoscopic'', in the sense that it fixes a particular coarse-graining length scale, $l$, and therefore excludes the recently discovered anomalous transport phenomena in the quantum spin-$1/2$ XXZ chain\cite{SF,Agrawal}, that emerge from linear fluctuating hydrodynamics in the limit of infinite coarse-graining length, $l \to \infty$.

\paragraph{Gauge modes in the hydrodynamics of isotropic integrable systems.} Given that nonlinear fluctuating hydrodynamics predicts purely diffusive fluctuations, the numerical discovery of Kardar-Parisi-Zhang universality in isotropic integrable systems appears puzzling. The resolution to this paradox is as follows: the Bethe-Boltzmann equation, Eq. \eqref{bb}, is inequivalent to the generalized hydrodynamics of infinitely many conservation laws. In particular, it is a scalar equation and contains limited information about the local direction of the magnetization. This fact was previously noted in the gapped phase of the spin-$1/2$ XXZ model\cite{Piroli2017}, but its implications at the isotropic point have not been addressed.

To clarify this issue, let us consider the spin-$1/2$ Heisenberg Hamiltonian
\begin{equation}
H = -J \sum_{i=1}^N \mathbf{S}_i \cdot \mathbf{S}_{i+1} - 2 \mathbf{h} \cdot \sum_{i=1}^N \mathbf{S}_i.
\end{equation}
In the absence of an applied magnetic field, $\mathbf{h}=0$, Bethe's solution to this model\cite{TakTBA} exhibits an $SU(2)$ gauge symmetry, due to the arbitrariness in the direction of the Bethe pseudovacuum, $\mathbf{\Omega} \in S^2$ (henceforth we neglect the $U(1)$ phase degree of freedom). This gauge symmetry extends to the thermodynamic Bethe ansatz (TBA) equations for the model. By contrast, when a magnetic field $\mathbf{h} \neq 0$ is applied, the rotational symmetry is broken and only pseudovacuum directions $\mathbf{\Omega} \parallel \mathbf{h}$ parallel or antiparallel to the applied field give rise to a set of Bethe eigenstates. For thermal states with applied magnetization $\mathbf{h}$, thermodynamic Bethe ansatz predicts a thermal expectation value
\begin{equation}
\mathbf{S} = \hat{\mathbf{h}} \left[ \frac{1}{2} - \sum_{n=1}^\infty \int_{-\infty}^{\infty} dk \, n\rho^n_k\right]
\end{equation}
for the spin density (here, $\rho^n_k$ denotes the density of occupied string modes with length $n$ and pseudomomentum $k$).

The crucial point is that in a hydrodynamic cell $\Gamma_i$, formation of an instantaneous, local magnetization $\mathbf{h}_{i}$ spontaneously breaks the pseudovacuum gauge symmetry. A full specification of the thermal state in $\Gamma_i$ therefore includes both the root densities, $\{\rho^n_k\}_i$ and the pseudovacuum vector $\mathbf{\Omega}_i = \mathbf{h}_i / \| \mathbf{h}_i \|$. The dynamics of the root densities is captured by the Bethe-Boltzmann equation, while the dynamics of the pseudovacuum is not.

The consequences of this simple observation are quite surprising. For example, it implies the existence of hydrodynamic states with slow spin dynamics and no quasiparticle dynamics. Consider a spin-$1/2$ Heisenberg chain, that has been coarse grained into hydrodynamic cells $\Gamma_i$ of length $l$. A perfectly admissible class of hydrodynamic states in this system have constant quasiparticle occupation numbers, $\{\rho^n_k\}_i = \{\rho^n_k\}$, yielding a constant absolute value of spin density $S = \frac{1}{2} - \sum_{n=1}^\infty \int_{-\infty}^{\infty} dk \, n \rho^n_k > 0$, and a local pseudovacuum $\mathbf{\Omega}_i$ that varies slowly with $i$, on a length scale $l_\mathbf{\Omega} \gg l$. (Here, $l$ should be regarded as the length scale of local equilibration for the short-wavelength quasiparticle degrees of freedom, which finite-temperature numerical simulations in the spin-$1/2$ XXZ chain\cite{Solv} suggest is of the order of the lattice scale.) Such states exhibit no time dynamics in the quasiparticle sector, which is encoded by the Bethe-Boltzmann equation Eq. \eqref{bb}. Meanwhile, slow modulations of the pseudovacuum are mean-field states by definition, whose long-wavelength, Hamiltonian dynamics satisfies the Landau-Lifshitz equation
\begin{equation}
\label{LL}
\partial_t \mathbf{\Omega} = \lambda \mathbf{\Omega} \times \partial_x^2 \mathbf{\Omega},
\end{equation}
where $\lambda = J/2$. By separation of scales, the microscopic quasiparticle degrees of freedom couple to $\mathbf{\Omega}$ as a local bath.

This class of hydrodynamic states includes the tunable $z$-polarized domain walls $\rho \propto (1+\mu \sigma_z)^{N/2} \otimes (1-\mu \sigma_z)^{N/2}$ that have frequently been considered in the literature\cite{Gobert,Bertini2016,KPZ3,Misguich,DW,Subdiff,Stephan,LL1,LL2}. In the continuum limit $N \to \infty$, the hydrodynamic state is determined as above, leading to a uniform initial condition for the quasiparticle mode occupancies and a domain wall initial condition for the pseudovacuum gauge field. From our perspective, the observed qualitative and quantitative similarities between the time-evolution of fully polarized domain walls in the isotropic spin-$1/2$ Heisenberg and Landau-Lifshitz models\cite{LL1,LL2}, constitute a test of the mean-field equation of motion, Eq. \eqref{LL}, at zero temperature.

These long-wavelength excitations can be interpreted as ``Goldstone modes'' of the quasiparticle hydrodynamics; they arise due to a redundancy in the specification of thermal states by scalar observables and transport negligible energy compared to the quasiparticle degrees of freedom. We expect precisely the same modes to arise for isotropic classically integrable magnets, since all local conserved charges generated by the transfer matrix approach are scalar\cite{Faddeev}. It is important to stress that a clean separation between ``gauge'' and ``quasiparticle'' degrees of freedom is an approximation that is contingent on a separation of scales. Since a full hydrodynamic description in the regime $l_{\mathbf{\Omega}} \sim l$ would require a more sophisticated microscopic understanding of the gauge dynamics than exists at present, we shall leave this question aside for now and focus on the regime $l_{\mathbf{\Omega}} \gg l$. The resulting physical picture, of short-wavelength quasiparticles furnishing a local bath for long-wavelength gauge modes, is very similar to the mechanism that generates KPZ physics in one-dimensional Bose gases\cite{Gangardt}. We now argue that this picture is sufficient to explain the emergence of KPZ scaling in isotropic integrable magnets.

\paragraph{Coarse-grained dynamics of soft gauge modes.} Let us consider the effect of coupling the Landau-Lifshitz equation to a local bath. Despite integrability of Eq. \eqref{LL} as written, the coarse-grained evolution it describes is not integrable, because lattice corrections to the effective description are ``dangerously irrelevant'' and alter the university class of the dynamics\cite{VasseurMoore}. (Precisely the same relation holds between the continuum Gross-Pitaevskii equation and its discretization\cite{Kulk2}.)

Obtaining the conserved modes of the Landau-Lifshitz evolution in a gauge-invariant manner is subtle, and standard parameterizations of the spin direction $\mathbf{\Omega}$ are ineffective. An elegant solution is to regard $\mathbf{\Omega}$ as the tangent vector of a space curve, with arc-length parameterized by position\cite{Lakshmanan}. Then the two gauge-invariant conserved modes are found to be energy density $\mathcal{E} = \kappa^2/2$ and torsion $\tau$ (geometrically, $\kappa$ and $\tau$ denote the curvature and torsion in the Frenet-Serret frame of the curve traced out by $\mathbf{\Omega}$). Discarding a dispersive term, this yields the coupled system of conservation laws

\begin{align}
\partial_t \mathcal{E} + \partial_x [\lambda(2\mathcal{E}\tau)] &= 0, \\
\partial_t \tau + \partial_x [\lambda(\tau^2 - \mathcal{E})] &= 0.
\end{align}

Na{\"i}vely, these equations represent the Euler-scale hydrodynamics of the Landau-Lifshitz evolution. However, their linearized dynamics is generically unstable. Physically, this pathology can be traced back to the fact that the energy carried by a soft mode is negligible at long length scales. More precisely, in a given stationary state which varies on the length scale, $l_{\mathbf{\Omega}}$, the total energy in a fluid cell of length $\mathcal{O}( l_{\mathbf{\Omega}})$ scales subextensively, as $\mathcal{O}( l_{\mathbf{\Omega}}^{-1})$. This indicates that deviations of $\mathcal{E}$ from zero are irrelevant in a renormalization group sense, so that in the Euler scaling limit, $\mathcal{E}=0$ and hydrodynamic instability is avoided. The coarse-grained time evolution of the residual torsional mode satisfies the stochastic Burgers equation
\begin{equation}
\label{torsion}
\partial_t \tau + \partial_x (\lambda \tau^2- D\partial_x \tau + \sigma \zeta_\tau) = 0,
\end{equation}
where $\zeta_\tau$ is a zero-mean white noise with $\quad \mathbb{E}[\zeta_\tau(x,t) \zeta_{\tau}(x',t')] = \delta(x-x')\delta(t-t')$, and equilibrium correlations $\chi = \langle \tau \tau \rangle$ of the torsional mode are constrained by the fluctuation-dissipation relation $\chi = \sigma^2/2D$. It follows that the height function $\eta$, defined by $\tau=\partial_x \eta$ satisfies the Kardar-Parisi-Zhang equation, and that the correlation functions of the torsional mode have superdiffusive scaling form\cite{SpohnFluct} $C(x,t) = \mathbb{E}[\tau(x,t) \tau(0,0)] = \chi f_{KPZ}(x/(\Gamma t)^{3/2})/(\Gamma t)^{3/2}$, where $\Gamma = 2\sqrt{2} \lambda$. Since in our parameterization\footnote{Stationary states with characteristic length scale $l_{\mathbf{\Omega}}$ and average spin $\mu/2$ along a given axis are helices with $\tau =\mu/l_\mathbf{\Omega}$ and $\mathcal{E} = (1-\mu^2)/2l_{\mathbf{\Omega}}^2$.}, the local magnetization density $\mu \propto \tau$, the emergence of KPZ scaling in isotropic one-dimensional magnets is explained.

We note that $\lambda$ will generically be renormalized from its bare value in the approximations leading to Eq. \eqref{torsion}, and predicting its precise value in specific models is beyond the scope of this work. Nevertheless, for both the quantum spin-$1/2$ Heisenberg chain and the classical integrable lattice Landau-Lifshitz model, our treatment of the pseudovacuum dynamics yields the coupling strength $\lambda/J = 0.5$, well within the expected margin of error\cite{SpohnFluct} of the numerically observed values\cite{KPZ4,KPZ5b}, $\lambda/J \approx 0.65$ and $\lambda/J \approx 0.68$ respectively. The fact that $\mathbf{\Omega}$ is a soft mode further explains a puzzling feature of this KPZ physics that was previously noted in the literature, namely its insensitivity to local energy conservation\cite{KPZ4}.

One novel prediction from our model, that might be tested numerically in quantum or classical integrable models, is the following: for time-evolution from weak domain walls\cite{KPZ3} with magnetization below the noise scale $0 < \mu < \sigma_{\mu}$ and orientation differing by an angle $\varphi$, the amplitude of the superdiffusive torsional mode will vary as $A \sim |\sin{(\varphi/2)}|$. Our model also predicts a crossover to a regime $\sigma_{\mu} < \mu$ in which the torsional mode becomes ballistic, with a propagation speed varying linearly in the initial magnetization, as $v_{ball} \sim \mu$. This result was obtained from generalized hydrodynamics in the spin-$1/2$ Heisenberg chain\cite{SF}; here we argue that the same phenomenon will arise in all quantum and classical integrable magnets with isotropic rotational symmetry. Indeed, this may explain the slow ballistic modes observed in recent classical simulations\cite{KPZ7}.

\paragraph{Discussion.} We have shown that a redundancy of description in the thermal states of isotropic integrable models, corresponding to the choice of Bethe pseudovacuum, gives rise to soft modes of the local magnetization that provide a channel for superdiffusive spin transport. We have further demonstrated that the coarse-grained time-evolution of these modes reduces to a single, torsional degree of freedom that satisfies the stochastic Burgers equation, explaining the numerical observation of KPZ scaling in such models\cite{KPZ1,KPZ2,KPZ3,KPZ4,KPZ5,KPZ5b,KPZ6,KPZ7}.

We now discuss why our theory implies that both integrability and isotropic symmetry are necessary to give rise to KPZ universality, in agreement with the most recent numerical findings\cite{KPZ6,KPZ7}. First, consider integrable systems. The vanishing Hessian in the nonlinear fluctuating hydrodynamics of quasiparticle modes indicates that in the absence of internal symmetries, corrections to ballistic dynamics will generically be diffusive\cite{DoyonFluct,KPZ5}. Thus among integrable many-particle systems, we expect that only those with a continuous symmetry of their thermal states, as specified by scalar observables, can exhibit KPZ physics. This class certainly includes isotropic integrable magnets. 
Our arguments should also extend directly to higher-dimensional internal symmetry groups than $G=SU(2)$, although the mode-coupling theory in the general case will be more complicated than Eq. \eqref{torsion}. To generate a freely propagating gauge mode, it seems necessary that $\textrm{dim}(G) \geq 2$; for example, in the Lieb-Liniger Bose gas, whose quantum fields possess a $U(1)$ phase symmetry, mean-field phase dynamics is coupled to mean-field density dynamics. 

We next turn to non-integrable magnets with isotropic symmetry. Generically, these will have four hydrodynamic modes, the energy $E$ and the three components of average spin, $\langle \mathbf{S} \rangle$ (note that $|\langle\mathbf{S}\rangle| \neq \langle |\mathbf{S}| \rangle$ on a fluid cell). One might argue that a separation of scales $l_{\mathbf{\Omega}}\gg l$ will again give rise to a KPZ torsional mode. In fact, treating the dynamics of $\mathbf{\Omega}$ and $S$ independently, as was necessary to obtain the torsional mode, is only consistent in integrable magnets. To see this, suppose we initialize an isotropic magnet in a long-wavelength excitation of $\mathbf{\Omega}$, at non-zero temperature. As the system equilibrates, gauge modes at the length scale $l_{\mathbf{\Omega}}$ excite scalar degrees of freedom, leading to a flow $l \to l_{\mathbf{\Omega}}$ of length scales. However, for non-integrable magnets, there are two scalar hydrodynamic modes $\{S,E\}$ per fluid cell, whose variance $\sigma^2 \sim l^{-1}$, while for integrable magnets, there is an extensive family $\{S,E,Q_2,\ldots,Q_n\}_{n\sim l}$ of such modes, all of which fluctuate diffusively with total variance $\sigma^2 \sim l^0$. It follows that as the magnet equilibrates, the dynamics of $S$ in a non-integrable system becomes deterministic and recouples to that of $\mathbf{\Omega}$, while in an integrable system, $S$ continues to fluctuate as part of a quasiparticle bath. This explains the observation of superdiffusive scaling of spin autocorrelation functions in isotropic, non-integrable magnets at short times\cite{KPZ5}, and also why it should cross over to the nonlinear fluctuating hydrodynamics prediction of diffusive scaling at long times\cite{KPZ6,ClassXXX,ClassXXZ}. 

Our discussion illustrates that the observation of KPZ physics in quantum integrable systems is essentially a classical phenomenon. We emphasize that the emergent classical degrees of freedom in our theory arise from the assumption of local equilibrium, rather than a low-temperature semiclassical description. An important question for future work is understanding how far the gauge modes that we identify can be described in the language of Bethe quasiparticles. Presumably, they are encoded by large strings of magnons\cite{DivD,GV}, but the correct order of limits is unclear. Recent analytical results on current operators in integrable, finite, spin chains\cite{PG1,PG2} might clarify this issue.

To summarize, our arguments explain a diverse set of numerical observations in one-dimensional quantum and classical magnets that were previously beyond theoretical techniques. We illustrate how Kardar-Parisi-Zhang physics can arise in isotropic integrable magnets based only on symmetry principles and model-independent features of the thermodynamics, providing a theoretical framework that is broad enough to encompass the seemingly universal nature of this phenomenon. Extending these ideas to a systematic understanding of the interplay between internal Lie group symmetries and emergent hydrodynamic behaviours in one-dimensional physical systems is a tantalizing goal for future investigation.

\textit{Acknowledgements ---} We thank A. G. Abanov, E. Altman, J. de Nardis, B. Doyon, M. Dupont, D. M. Gangardt, S. Gopalakrishnan, E. Ilievski, J. E. Moore, B. Pozsgay, H. Spohn and R. Vasseur for illuminating discussions, and X. Cao and C. Karrasch for collaborations on related work. We are grateful to the organizers and hosts of the workshop ``Emergent Hydrodynamics in Low Dimensional Quantum Systems'' at the International Institute of Physics in Natal, Brazil, and to the organizers and hosts of the summer school ``Dynamics and Disorder in Quantum Many Body Systems Far-from-Equilibrium'' in Les Houches, France. This research was supported in part by the International Centre for Theoretical Sciences in Bangalore, India, during a visit for the program ``Thermalization, Many Body Localization and Hydrodynamics'' (code: ICTS/hydrodynamics2019/11). We acknowledge support from the the DRINQS program of the Defense Advanced Research Projects Agency.
\bibliography{softbib}

\begin{thebibliography}{59}%
\makeatletter
\providecommand \@ifxundefined [1]{%
 \@ifx{#1\undefined}
}%
\providecommand \@ifnum [1]{%
 \ifnum #1\expandafter \@firstoftwo
 \else \expandafter \@secondoftwo
 \fi
}%
\providecommand \@ifx [1]{%
 \ifx #1\expandafter \@firstoftwo
 \else \expandafter \@secondoftwo
 \fi
}%
\providecommand \natexlab [1]{#1}%
\providecommand \enquote  [1]{``#1''}%
\providecommand \bibnamefont  [1]{#1}%
\providecommand \bibfnamefont [1]{#1}%
\providecommand \citenamefont [1]{#1}%
\providecommand \href@noop [0]{\@secondoftwo}%
\providecommand \href [0]{\begingroup \@sanitize@url \@href}%
\providecommand \@href[1]{\@@startlink{#1}\@@href}%
\providecommand \@@href[1]{\endgroup#1\@@endlink}%
\providecommand \@sanitize@url [0]{\catcode `\\12\catcode `\$12\catcode
  `\&12\catcode `\#12\catcode `\^12\catcode `\_12\catcode `\%12\relax}%
\providecommand \@@startlink[1]{}%
\providecommand \@@endlink[0]{}%
\providecommand \url  [0]{\begingroup\@sanitize@url \@url }%
\providecommand \@url [1]{\endgroup\@href {#1}{\urlprefix }}%
\providecommand \urlprefix  [0]{URL }%
\providecommand \Eprint [0]{\href }%
\providecommand \doibase [0]{http://dx.doi.org/}%
\providecommand \selectlanguage [0]{\@gobble}%
\providecommand \bibinfo  [0]{\@secondoftwo}%
\providecommand \bibfield  [0]{\@secondoftwo}%
\providecommand \translation [1]{[#1]}%
\providecommand \BibitemOpen [0]{}%
\providecommand \bibitemStop [0]{}%
\providecommand \bibitemNoStop [0]{.\EOS\space}%
\providecommand \EOS [0]{\spacefactor3000\relax}%
\providecommand \BibitemShut  [1]{\csname bibitem#1\endcsname}%
\let\auto@bib@innerbib\@empty
\bibitem [{\citenamefont {Alder}\ and\ \citenamefont {Wainwright}(1970)}]{AW}%
  \BibitemOpen
  \bibfield  {author} {\bibinfo {author} {\bibfnamefont {B.~J.}\ \bibnamefont
  {Alder}}\ and\ \bibinfo {author} {\bibfnamefont {T.~E.}\ \bibnamefont
  {Wainwright}},\ }\href {\doibase 10.1103/PhysRevA.1.18} {\bibfield  {journal}
  {\bibinfo  {journal} {Phys. Rev. A}\ }\textbf {\bibinfo {volume} {1}},\
  \bibinfo {pages} {18} (\bibinfo {year} {1970})}\BibitemShut {NoStop}%
\bibitem [{\citenamefont {van Beijeren}(2012)}]{vanB}%
  \BibitemOpen
  \bibfield  {author} {\bibinfo {author} {\bibfnamefont {H.}~\bibnamefont {van
  Beijeren}},\ }\href {\doibase 10.1103/PhysRevLett.108.180601} {\bibfield
  {journal} {\bibinfo  {journal} {Phys. Rev. Lett.}\ }\textbf {\bibinfo
  {volume} {108}},\ \bibinfo {pages} {180601} (\bibinfo {year}
  {2012})}\BibitemShut {NoStop}%
\bibitem [{\citenamefont {Spohn}(2016)}]{SpohnFluct}%
  \BibitemOpen
  \bibfield  {author} {\bibinfo {author} {\bibfnamefont {H.}~\bibnamefont
  {Spohn}},\ }\enquote {\bibinfo {title} {Fluctuating hydrodynamics approach to
  equilibrium time correlations for anharmonic chains},}\ in\ \href {\doibase
  10.1007/978-3-319-29261-8_3} {\emph {\bibinfo {booktitle} {Thermal Transport
  in Low Dimensions}}},\ \bibinfo {editor} {edited by\ \bibinfo {editor}
  {\bibfnamefont {S.}~\bibnamefont {Lepri}}}\ (\bibinfo  {publisher} {Springer
  International Publishing},\ \bibinfo {address} {Cham},\ \bibinfo {year}
  {2016})\ pp.\ \bibinfo {pages} {107--158}\BibitemShut {NoStop}%
\bibitem [{\citenamefont {Kulkarni}\ and\ \citenamefont
  {Lamacraft}(2013)}]{KulkLam}%
  \BibitemOpen
  \bibfield  {author} {\bibinfo {author} {\bibfnamefont {M.}~\bibnamefont
  {Kulkarni}}\ and\ \bibinfo {author} {\bibfnamefont {A.}~\bibnamefont
  {Lamacraft}},\ }\href {\doibase 10.1103/PhysRevA.88.021603} {\bibfield
  {journal} {\bibinfo  {journal} {Phys. Rev. A}\ }\textbf {\bibinfo {volume}
  {88}},\ \bibinfo {pages} {021603} (\bibinfo {year} {2013})}\BibitemShut
  {NoStop}%
\bibitem [{\citenamefont {Popkov}\ \emph {et~al.}(2015)\citenamefont {Popkov},
  \citenamefont {Schadschneider}, \citenamefont {Schmidt},\ and\ \citenamefont
  {Sch{\"u}tz}}]{Fib}%
  \BibitemOpen
  \bibfield  {author} {\bibinfo {author} {\bibfnamefont {V.}~\bibnamefont
  {Popkov}}, \bibinfo {author} {\bibfnamefont {A.}~\bibnamefont
  {Schadschneider}}, \bibinfo {author} {\bibfnamefont {J.}~\bibnamefont
  {Schmidt}}, \ and\ \bibinfo {author} {\bibfnamefont {G.~M.}\ \bibnamefont
  {Sch{\"u}tz}},\ }\href {\doibase 10.1073/pnas.1512261112} {\bibfield
  {journal} {\bibinfo  {journal} {Proc. Natl. Acad. Sci.}\ }\textbf {\bibinfo
  {volume} {112}},\ \bibinfo {pages} {12645} (\bibinfo {year}
  {2015})}\BibitemShut {NoStop}%
\bibitem [{\citenamefont {Rigol}\ \emph {et~al.}(2008)\citenamefont {Rigol},
  \citenamefont {Dunjko},\ and\ \citenamefont {Olshanii}}]{Rigol}%
  \BibitemOpen
  \bibfield  {author} {\bibinfo {author} {\bibfnamefont {M.}~\bibnamefont
  {Rigol}}, \bibinfo {author} {\bibfnamefont {V.}~\bibnamefont {Dunjko}}, \
  and\ \bibinfo {author} {\bibfnamefont {M.}~\bibnamefont {Olshanii}},\ }\href
  {https://doi.org/10.1038/nature06838 http://10.0.4.14/nature06838
  https://www.nature.com/articles/nature06838{\#}supplementary-information}
  {\bibfield  {journal} {\bibinfo  {journal} {Nature}\ }\textbf {\bibinfo
  {volume} {452}},\ \bibinfo {pages} {854} (\bibinfo {year}
  {2008})}\BibitemShut {NoStop}%
\bibitem [{\citenamefont {Barthel}\ and\ \citenamefont
  {Schollw\"ock}(2008)}]{BS}%
  \BibitemOpen
  \bibfield  {author} {\bibinfo {author} {\bibfnamefont {T.}~\bibnamefont
  {Barthel}}\ and\ \bibinfo {author} {\bibfnamefont {U.}~\bibnamefont
  {Schollw\"ock}},\ }\href {\doibase 10.1103/PhysRevLett.100.100601} {\bibfield
   {journal} {\bibinfo  {journal} {Phys. Rev. Lett.}\ }\textbf {\bibinfo
  {volume} {100}},\ \bibinfo {pages} {100601} (\bibinfo {year}
  {2008})}\BibitemShut {NoStop}%
\bibitem [{\citenamefont {Percus}(1969)}]{Percus}%
  \BibitemOpen
  \bibfield  {author} {\bibinfo {author} {\bibfnamefont {J.~K.}\ \bibnamefont
  {Percus}},\ }\href {\doibase 10.1063/1.1692711} {\bibfield  {journal}
  {\bibinfo  {journal} {The Physics of Fluids}\ }\textbf {\bibinfo {volume}
  {12}},\ \bibinfo {pages} {1560} (\bibinfo {year} {1969})}\BibitemShut
  {NoStop}%
\bibitem [{\citenamefont {{Zakharov}}(1971)}]{Zakharov}%
  \BibitemOpen
  \bibfield  {author} {\bibinfo {author} {\bibfnamefont {V.~E.}\ \bibnamefont
  {{Zakharov}}},\ }\href@noop {} {\bibfield  {journal} {\bibinfo  {journal} {J.
  Exp. Theor. Phys.}\ }\textbf {\bibinfo {volume} {33}},\ \bibinfo {pages}
  {538} (\bibinfo {year} {1971})}\BibitemShut {NoStop}%
\bibitem [{\citenamefont {Boldrighini}\ \emph {et~al.}(1983)\citenamefont
  {Boldrighini}, \citenamefont {Dobrushin},\ and\ \citenamefont
  {Sukhov}}]{Boldrighini}%
  \BibitemOpen
  \bibfield  {author} {\bibinfo {author} {\bibfnamefont {C.}~\bibnamefont
  {Boldrighini}}, \bibinfo {author} {\bibfnamefont {R.~L.}\ \bibnamefont
  {Dobrushin}}, \ and\ \bibinfo {author} {\bibfnamefont {Y.~M.}\ \bibnamefont
  {Sukhov}},\ }\href {\doibase 10.1007/BF01019499} {\bibfield  {journal}
  {\bibinfo  {journal} {Journal of Statistical Physics}\ }\textbf {\bibinfo
  {volume} {31}},\ \bibinfo {pages} {577} (\bibinfo {year} {1983})}\BibitemShut
  {NoStop}%
\bibitem [{\citenamefont {El}\ and\ \citenamefont {Kamchatnov}(2005)}]{EK}%
  \BibitemOpen
  \bibfield  {author} {\bibinfo {author} {\bibfnamefont {G.~A.}\ \bibnamefont
  {El}}\ and\ \bibinfo {author} {\bibfnamefont {A.~M.}\ \bibnamefont
  {Kamchatnov}},\ }\href {\doibase 10.1103/PhysRevLett.95.204101} {\bibfield
  {journal} {\bibinfo  {journal} {Phys. Rev. Lett.}\ }\textbf {\bibinfo
  {volume} {95}},\ \bibinfo {pages} {204101} (\bibinfo {year}
  {2005})}\BibitemShut {NoStop}%
\bibitem [{\citenamefont {Bertini}\ \emph {et~al.}(2016)\citenamefont
  {Bertini}, \citenamefont {Collura}, \citenamefont {De~Nardis},\ and\
  \citenamefont {Fagotti}}]{Bertini2016}%
  \BibitemOpen
  \bibfield  {author} {\bibinfo {author} {\bibfnamefont {B.}~\bibnamefont
  {Bertini}}, \bibinfo {author} {\bibfnamefont {M.}~\bibnamefont {Collura}},
  \bibinfo {author} {\bibfnamefont {J.}~\bibnamefont {De~Nardis}}, \ and\
  \bibinfo {author} {\bibfnamefont {M.}~\bibnamefont {Fagotti}},\ }\href
  {\doibase 10.1103/PhysRevLett.117.207201} {\bibfield  {journal} {\bibinfo
  {journal} {Phys. Rev. Lett.}\ }\textbf {\bibinfo {volume} {117}},\ \bibinfo
  {pages} {207201} (\bibinfo {year} {2016})}\BibitemShut {NoStop}%
\bibitem [{\citenamefont {Castro-Alvaredo}\ \emph {et~al.}(2016)\citenamefont
  {Castro-Alvaredo}, \citenamefont {Doyon},\ and\ \citenamefont
  {Yoshimura}}]{Castro-Alvaredo2016}%
  \BibitemOpen
  \bibfield  {author} {\bibinfo {author} {\bibfnamefont {O.~A.}\ \bibnamefont
  {Castro-Alvaredo}}, \bibinfo {author} {\bibfnamefont {B.}~\bibnamefont
  {Doyon}}, \ and\ \bibinfo {author} {\bibfnamefont {T.}~\bibnamefont
  {Yoshimura}},\ }\href {\doibase 10.1103/PhysRevX.6.041065} {\bibfield
  {journal} {\bibinfo  {journal} {Phys. Rev. X}\ }\textbf {\bibinfo {volume}
  {6}},\ \bibinfo {pages} {041065} (\bibinfo {year} {2016})}\BibitemShut
  {NoStop}%
\bibitem [{\citenamefont {Bulchandani}\ \emph {et~al.}(2018)\citenamefont
  {Bulchandani}, \citenamefont {Vasseur}, \citenamefont {Karrasch},\ and\
  \citenamefont {Moore}}]{Bulchandani2018}%
  \BibitemOpen
  \bibfield  {author} {\bibinfo {author} {\bibfnamefont {V.~B.}\ \bibnamefont
  {Bulchandani}}, \bibinfo {author} {\bibfnamefont {R.}~\bibnamefont
  {Vasseur}}, \bibinfo {author} {\bibfnamefont {C.}~\bibnamefont {Karrasch}}, \
  and\ \bibinfo {author} {\bibfnamefont {J.~E.}\ \bibnamefont {Moore}},\ }\href
  {\doibase 10.1103/PhysRevB.97.045407} {\bibfield  {journal} {\bibinfo
  {journal} {Phys. Rev. B}\ }\textbf {\bibinfo {volume} {97}},\ \bibinfo
  {pages} {045407} (\bibinfo {year} {2018})}\BibitemShut {NoStop}%
\bibitem [{\citenamefont {Schemmer}\ \emph {et~al.}(2019)\citenamefont
  {Schemmer}, \citenamefont {Bouchoule}, \citenamefont {Doyon},\ and\
  \citenamefont {Dubail}}]{Schemmer}%
  \BibitemOpen
  \bibfield  {author} {\bibinfo {author} {\bibfnamefont {M.}~\bibnamefont
  {Schemmer}}, \bibinfo {author} {\bibfnamefont {I.}~\bibnamefont {Bouchoule}},
  \bibinfo {author} {\bibfnamefont {B.}~\bibnamefont {Doyon}}, \ and\ \bibinfo
  {author} {\bibfnamefont {J.}~\bibnamefont {Dubail}},\ }\href {\doibase
  10.1103/PhysRevLett.122.090601} {\bibfield  {journal} {\bibinfo  {journal}
  {Phys. Rev. Lett.}\ }\textbf {\bibinfo {volume} {122}},\ \bibinfo {pages}
  {090601} (\bibinfo {year} {2019})}\BibitemShut {NoStop}%
\bibitem [{\citenamefont {\ifmmode \check{Z}\else
  \v{Z}\fi{}nidari\ifmmode~\check{c}\else \v{c}\fi{}}(2011)}]{KPZ1}%
  \BibitemOpen
  \bibfield  {author} {\bibinfo {author} {\bibfnamefont {M.}~\bibnamefont
  {\ifmmode \check{Z}\else \v{Z}\fi{}nidari\ifmmode~\check{c}\else
  \v{c}\fi{}}},\ }\href {\doibase 10.1103/PhysRevLett.106.220601} {\bibfield
  {journal} {\bibinfo  {journal} {Phys. Rev. Lett.}\ }\textbf {\bibinfo
  {volume} {106}},\ \bibinfo {pages} {220601} (\bibinfo {year}
  {2011})}\BibitemShut {NoStop}%
\bibitem [{\citenamefont {Prosen}\ and\ \citenamefont {\ifmmode \check{Z}\else
  \v{Z}\fi{}unkovi\ifmmode~\check{c}\else \v{c}\fi{}}(2013)}]{KPZ2}%
  \BibitemOpen
  \bibfield  {author} {\bibinfo {author} {\bibfnamefont {T.}~\bibnamefont
  {Prosen}}\ and\ \bibinfo {author} {\bibfnamefont {B.}~\bibnamefont {\ifmmode
  \check{Z}\else \v{Z}\fi{}unkovi\ifmmode~\check{c}\else \v{c}\fi{}}},\ }\href
  {\doibase 10.1103/PhysRevLett.111.040602} {\bibfield  {journal} {\bibinfo
  {journal} {Phys. Rev. Lett.}\ }\textbf {\bibinfo {volume} {111}},\ \bibinfo
  {pages} {040602} (\bibinfo {year} {2013})}\BibitemShut {NoStop}%
\bibitem [{\citenamefont {Ljubotina}\ \emph {et~al.}(2017)\citenamefont
  {Ljubotina}, \citenamefont {{\v{Z}}nidari\ifmmode~\check{c}\else
  \v{c}\fi{}},\ and\ \citenamefont {Prosen}}]{KPZ3}%
  \BibitemOpen
  \bibfield  {author} {\bibinfo {author} {\bibfnamefont {M.}~\bibnamefont
  {Ljubotina}}, \bibinfo {author} {\bibfnamefont {M.}~\bibnamefont
  {{\v{Z}}nidari\ifmmode~\check{c}\else \v{c}\fi{}}}, \ and\ \bibinfo {author}
  {\bibfnamefont {T.}~\bibnamefont {Prosen}},\ }\href {\doibase
  10.1038/ncomms16117} {\bibfield  {journal} {\bibinfo  {journal} {Nature
  Communications}\ }\textbf {\bibinfo {volume} {8}} (\bibinfo {year} {2017}),\
  10.1038/ncomms16117}\BibitemShut {NoStop}%
\bibitem [{\citenamefont {Ljubotina}\ \emph {et~al.}(2019)\citenamefont
  {Ljubotina}, \citenamefont {\ifmmode \check{Z}\else
  \v{Z}\fi{}nidari\ifmmode~\check{c}\else \v{c}\fi{}},\ and\ \citenamefont
  {Prosen}}]{KPZ4}%
  \BibitemOpen
  \bibfield  {author} {\bibinfo {author} {\bibfnamefont {M.}~\bibnamefont
  {Ljubotina}}, \bibinfo {author} {\bibfnamefont {M.}~\bibnamefont {\ifmmode
  \check{Z}\else \v{Z}\fi{}nidari\ifmmode~\check{c}\else \v{c}\fi{}}}, \ and\
  \bibinfo {author} {\bibfnamefont {T.}~\bibnamefont {Prosen}},\ }\href
  {\doibase 10.1103/PhysRevLett.122.210602} {\bibfield  {journal} {\bibinfo
  {journal} {Phys. Rev. Lett.}\ }\textbf {\bibinfo {volume} {122}},\ \bibinfo
  {pages} {210602} (\bibinfo {year} {2019})}\BibitemShut {NoStop}%
\bibitem [{\citenamefont {Nardis}\ \emph
  {et~al.}(2019{\natexlab{a}})\citenamefont {Nardis}, \citenamefont {Medenjak},
  \citenamefont {Karrasch},\ and\ \citenamefont {Ilievski}}]{KPZ5}%
  \BibitemOpen
  \bibfield  {author} {\bibinfo {author} {\bibfnamefont {J.~D.}\ \bibnamefont
  {Nardis}}, \bibinfo {author} {\bibfnamefont {M.}~\bibnamefont {Medenjak}},
  \bibinfo {author} {\bibfnamefont {C.}~\bibnamefont {Karrasch}}, \ and\
  \bibinfo {author} {\bibfnamefont {E.}~\bibnamefont {Ilievski}},\ }\href@noop
  {} {\  (\bibinfo {year} {2019}{\natexlab{a}})},\ \Eprint
  {http://arxiv.org/abs/1903.07598} {arXiv:1903.07598 [cond-mat.stat-mech]}
  \BibitemShut {NoStop}%
\bibitem [{\citenamefont {Das}\ \emph {et~al.}(2019)\citenamefont {Das},
  \citenamefont {Kulkarni}, \citenamefont {Spohn},\ and\ \citenamefont
  {Dhar}}]{KPZ5b}%
  \BibitemOpen
  \bibfield  {author} {\bibinfo {author} {\bibfnamefont {A.}~\bibnamefont
  {Das}}, \bibinfo {author} {\bibfnamefont {M.}~\bibnamefont {Kulkarni}},
  \bibinfo {author} {\bibfnamefont {H.}~\bibnamefont {Spohn}}, \ and\ \bibinfo
  {author} {\bibfnamefont {A.}~\bibnamefont {Dhar}},\ }\href {\doibase
  10.1103/PhysRevE.100.042116} {\bibfield  {journal} {\bibinfo  {journal}
  {Phys. Rev. E}\ }\textbf {\bibinfo {volume} {100}},\ \bibinfo {pages}
  {042116} (\bibinfo {year} {2019})}\BibitemShut {NoStop}%
\bibitem [{\citenamefont {Dupont}\ and\ \citenamefont {Moore}(2019)}]{KPZ6}%
  \BibitemOpen
  \bibfield  {author} {\bibinfo {author} {\bibfnamefont {M.}~\bibnamefont
  {Dupont}}\ and\ \bibinfo {author} {\bibfnamefont {J.~E.}\ \bibnamefont
  {Moore}},\ }\href@noop {} {\  (\bibinfo {year} {2019})},\ \Eprint
  {http://arxiv.org/abs/1907.12115} {arXiv:1907.12115 [cond-mat.str-el]}
  \BibitemShut {NoStop}%
\bibitem [{\citenamefont {Krajnik}\ and\ \citenamefont {Prosen}(2019)}]{KPZ7}%
  \BibitemOpen
  \bibfield  {author} {\bibinfo {author} {\bibfnamefont {Z.}~\bibnamefont
  {Krajnik}}\ and\ \bibinfo {author} {\bibfnamefont {T.}~\bibnamefont
  {Prosen}},\ }\href@noop {} {\  (\bibinfo {year} {2019})},\ \Eprint
  {http://arxiv.org/abs/1909.03799} {arXiv:1909.03799 [cond-mat.stat-mech]}
  \BibitemShut {NoStop}%
\bibitem [{\citenamefont {Doyon}\ and\ \citenamefont
  {Myers}(2019)}]{DoyonFluct}%
  \BibitemOpen
  \bibfield  {author} {\bibinfo {author} {\bibfnamefont {B.}~\bibnamefont
  {Doyon}}\ and\ \bibinfo {author} {\bibfnamefont {J.}~\bibnamefont {Myers}},\
  }\href@noop {} {\  (\bibinfo {year} {2019})},\ \Eprint
  {http://arxiv.org/abs/1902.00320} {arXiv:1902.00320 [cond-mat.stat-mech]}
  \BibitemShut {NoStop}%
\bibitem [{\citenamefont {Gopalakrishnan}\ and\ \citenamefont
  {Vasseur}(2019)}]{GV}%
  \BibitemOpen
  \bibfield  {author} {\bibinfo {author} {\bibfnamefont {S.}~\bibnamefont
  {Gopalakrishnan}}\ and\ \bibinfo {author} {\bibfnamefont {R.}~\bibnamefont
  {Vasseur}},\ }\href {\doibase 10.1103/PhysRevLett.122.127202} {\bibfield
  {journal} {\bibinfo  {journal} {Phys. Rev. Lett.}\ }\textbf {\bibinfo
  {volume} {122}},\ \bibinfo {pages} {127202} (\bibinfo {year}
  {2019})}\BibitemShut {NoStop}%
\bibitem [{\citenamefont {Fendley}\ and\ \citenamefont
  {Saleur}(1996)}]{FendleySaleur}%
  \BibitemOpen
  \bibfield  {author} {\bibinfo {author} {\bibfnamefont {P.}~\bibnamefont
  {Fendley}}\ and\ \bibinfo {author} {\bibfnamefont {H.}~\bibnamefont
  {Saleur}},\ }\href {\doibase 10.1103/PhysRevB.54.10845} {\bibfield  {journal}
  {\bibinfo  {journal} {Phys. Rev. B}\ }\textbf {\bibinfo {volume} {54}},\
  \bibinfo {pages} {10845} (\bibinfo {year} {1996})}\BibitemShut {NoStop}%
\bibitem [{\citenamefont {De~Nardis}\ \emph {et~al.}(2018)\citenamefont
  {De~Nardis}, \citenamefont {Bernard},\ and\ \citenamefont {Doyon}}]{Diff1}%
  \BibitemOpen
  \bibfield  {author} {\bibinfo {author} {\bibfnamefont {J.}~\bibnamefont
  {De~Nardis}}, \bibinfo {author} {\bibfnamefont {D.}~\bibnamefont {Bernard}},
  \ and\ \bibinfo {author} {\bibfnamefont {B.}~\bibnamefont {Doyon}},\ }\href
  {\doibase 10.1103/PhysRevLett.121.160603} {\bibfield  {journal} {\bibinfo
  {journal} {Phys. Rev. Lett.}\ }\textbf {\bibinfo {volume} {121}},\ \bibinfo
  {pages} {160603} (\bibinfo {year} {2018})}\BibitemShut {NoStop}%
\bibitem [{\citenamefont {Gopalakrishnan}\ \emph {et~al.}(2018)\citenamefont
  {Gopalakrishnan}, \citenamefont {Huse}, \citenamefont {Khemani},\ and\
  \citenamefont {Vasseur}}]{Diff2}%
  \BibitemOpen
  \bibfield  {author} {\bibinfo {author} {\bibfnamefont {S.}~\bibnamefont
  {Gopalakrishnan}}, \bibinfo {author} {\bibfnamefont {D.~A.}\ \bibnamefont
  {Huse}}, \bibinfo {author} {\bibfnamefont {V.}~\bibnamefont {Khemani}}, \
  and\ \bibinfo {author} {\bibfnamefont {R.}~\bibnamefont {Vasseur}},\ }\href
  {\doibase 10.1103/PhysRevB.98.220303} {\bibfield  {journal} {\bibinfo
  {journal} {Phys. Rev. B}\ }\textbf {\bibinfo {volume} {98}},\ \bibinfo
  {pages} {220303} (\bibinfo {year} {2018})}\BibitemShut {NoStop}%
\bibitem [{\citenamefont {Nardis}\ \emph
  {et~al.}(2019{\natexlab{b}})\citenamefont {Nardis}, \citenamefont {Bernard},\
  and\ \citenamefont {Doyon}}]{Diff3}%
  \BibitemOpen
  \bibfield  {author} {\bibinfo {author} {\bibfnamefont {J.~D.}\ \bibnamefont
  {Nardis}}, \bibinfo {author} {\bibfnamefont {D.}~\bibnamefont {Bernard}}, \
  and\ \bibinfo {author} {\bibfnamefont {B.}~\bibnamefont {Doyon}},\ }\href
  {\doibase 10.21468/SciPostPhys.6.4.049} {\bibfield  {journal} {\bibinfo
  {journal} {SciPost Phys.}\ }\textbf {\bibinfo {volume} {6}},\ \bibinfo
  {pages} {49} (\bibinfo {year} {2019}{\natexlab{b}})}\BibitemShut {NoStop}%
\bibitem [{\citenamefont {Boldrighini}\ and\ \citenamefont
  {Suhov}(1997)}]{Boldrighini2}%
  \BibitemOpen
  \bibfield  {author} {\bibinfo {author} {\bibfnamefont {C.}~\bibnamefont
  {Boldrighini}}\ and\ \bibinfo {author} {\bibfnamefont {Y.}~\bibnamefont
  {Suhov}},\ }\href {\doibase 10.1007/s002200050218} {\bibfield  {journal}
  {\bibinfo  {journal} {Communications in Mathematical Physics}\ }\textbf
  {\bibinfo {volume} {189}},\ \bibinfo {pages} {577} (\bibinfo {year}
  {1997})}\BibitemShut {NoStop}%
\bibitem [{\citenamefont {Doyon}\ and\ \citenamefont {Spohn}(2017)}]{DS}%
  \BibitemOpen
  \bibfield  {author} {\bibinfo {author} {\bibfnamefont {B.}~\bibnamefont
  {Doyon}}\ and\ \bibinfo {author} {\bibfnamefont {H.}~\bibnamefont {Spohn}},\
  }\href {\doibase 10.1088/1742-5468/aa7abf} {\bibfield  {journal} {\bibinfo
  {journal} {J. Stat. Mech.: Theory Exp}\ }\textbf {\bibinfo {volume} {2017}},\
  \bibinfo {pages} {073210} (\bibinfo {year} {2017})}\BibitemShut {NoStop}%
\bibitem [{\citenamefont {Cao}\ \emph {et~al.}(2018)\citenamefont {Cao},
  \citenamefont {Bulchandani},\ and\ \citenamefont {Moore}}]{Cao2018}%
  \BibitemOpen
  \bibfield  {author} {\bibinfo {author} {\bibfnamefont {X.}~\bibnamefont
  {Cao}}, \bibinfo {author} {\bibfnamefont {V.~B.}\ \bibnamefont
  {Bulchandani}}, \ and\ \bibinfo {author} {\bibfnamefont {J.~E.}\ \bibnamefont
  {Moore}},\ }\href {\doibase 10.1103/PhysRevLett.120.164101} {\bibfield
  {journal} {\bibinfo  {journal} {Phys. Rev. Lett.}\ }\textbf {\bibinfo
  {volume} {120}},\ \bibinfo {pages} {164101} (\bibinfo {year}
  {2018})}\BibitemShut {NoStop}%
\bibitem [{\citenamefont {Ilievski}\ \emph {et~al.}(2018)\citenamefont
  {Ilievski}, \citenamefont {De~Nardis}, \citenamefont {Medenjak},\ and\
  \citenamefont {Prosen}}]{DivD}%
  \BibitemOpen
  \bibfield  {author} {\bibinfo {author} {\bibfnamefont {E.}~\bibnamefont
  {Ilievski}}, \bibinfo {author} {\bibfnamefont {J.}~\bibnamefont {De~Nardis}},
  \bibinfo {author} {\bibfnamefont {M.}~\bibnamefont {Medenjak}}, \ and\
  \bibinfo {author} {\bibfnamefont {T.}~\bibnamefont {Prosen}},\ }\href
  {\doibase 10.1103/PhysRevLett.121.230602} {\bibfield  {journal} {\bibinfo
  {journal} {Phys. Rev. Lett.}\ }\textbf {\bibinfo {volume} {121}},\ \bibinfo
  {pages} {230602} (\bibinfo {year} {2018})}\BibitemShut {NoStop}%
\bibitem [{Note1()}]{Note1}%
  \BibitemOpen
  \bibinfo {note} {Using the operator identity $\protect \mathaccentV
  {hat}05E{M}^{-1} - \protect \mathaccentV {hat}05E{\theta } \protect
  \mathaccentV {hat}05E{\alpha } = \protect \mathaccentV {hat}05E{1}$, one can
  write this in the simplified form $G^k_{k'k''} = - \protect \frac {1}{2}
  \protect \frac {\lambda _k}{\rho ^t_k} \protect \frac {\alpha
  _{k'k''}}{\lambda _{k'}\lambda _{k''}}(v_{k'}-v_{k''})(\delta _{kk'} - \delta
  _{kk''})$, which matches a subsequent calculation\cite {Subs} of $G$, up to a
  choice of normalization in the definition of $C$.}\BibitemShut {Stop}%
\bibitem [{\citenamefont {El}\ \emph {et~al.}(2011)\citenamefont {El},
  \citenamefont {Kamchatnov}, \citenamefont {Pavlov},\ and\ \citenamefont
  {Zykov}}]{Int1}%
  \BibitemOpen
  \bibfield  {author} {\bibinfo {author} {\bibfnamefont {G.~A.}\ \bibnamefont
  {El}}, \bibinfo {author} {\bibfnamefont {A.~M.}\ \bibnamefont {Kamchatnov}},
  \bibinfo {author} {\bibfnamefont {M.~V.}\ \bibnamefont {Pavlov}}, \ and\
  \bibinfo {author} {\bibfnamefont {S.~A.}\ \bibnamefont {Zykov}},\ }\href
  {\doibase 10.1007/s00332-010-9080-z} {\bibfield  {journal} {\bibinfo
  {journal} {Journal of Nonlinear Science}\ }\textbf {\bibinfo {volume} {21}},\
  \bibinfo {pages} {151} (\bibinfo {year} {2011})}\BibitemShut {NoStop}%
\bibitem [{\citenamefont {Bulchandani}(2017)}]{Int2}%
  \BibitemOpen
  \bibfield  {author} {\bibinfo {author} {\bibfnamefont {V.~B.}\ \bibnamefont
  {Bulchandani}},\ }\href@noop {} {\bibfield  {journal} {\bibinfo  {journal}
  {J. Phys. A}\ }\textbf {\bibinfo {volume} {50}} (\bibinfo {year}
  {2017})}\BibitemShut {NoStop}%
\bibitem [{\citenamefont {Gopalakrishnan}\ \emph {et~al.}(2019)\citenamefont
  {Gopalakrishnan}, \citenamefont {Vasseur},\ and\ \citenamefont {Ware}}]{SF}%
  \BibitemOpen
  \bibfield  {author} {\bibinfo {author} {\bibfnamefont {S.}~\bibnamefont
  {Gopalakrishnan}}, \bibinfo {author} {\bibfnamefont {R.}~\bibnamefont
  {Vasseur}}, \ and\ \bibinfo {author} {\bibfnamefont {B.}~\bibnamefont
  {Ware}},\ }\href {\doibase 10.1073/pnas.1906914116} {\bibfield  {journal}
  {\bibinfo  {journal} {Proc. Natl. Acad. Sci.}\ }\textbf {\bibinfo {volume}
  {116}},\ \bibinfo {pages} {16250} (\bibinfo {year} {2019})}\BibitemShut
  {NoStop}%
\bibitem [{\citenamefont {Agrawal}\ \emph {et~al.}(2019)\citenamefont
  {Agrawal}, \citenamefont {Gopalakrishnan}, \citenamefont {Vasseur},\ and\
  \citenamefont {Ware}}]{Agrawal}%
  \BibitemOpen
  \bibfield  {author} {\bibinfo {author} {\bibfnamefont {U.}~\bibnamefont
  {Agrawal}}, \bibinfo {author} {\bibfnamefont {S.}~\bibnamefont
  {Gopalakrishnan}}, \bibinfo {author} {\bibfnamefont {R.}~\bibnamefont
  {Vasseur}}, \ and\ \bibinfo {author} {\bibfnamefont {B.}~\bibnamefont
  {Ware}},\ }\href@noop {} {\  (\bibinfo {year} {2019})},\ \Eprint
  {http://arxiv.org/abs/1909.05263} {arXiv:1909.05263 [cond-mat.stat-mech]}
  \BibitemShut {NoStop}%
\bibitem [{\citenamefont {Piroli}\ \emph {et~al.}(2017)\citenamefont {Piroli},
  \citenamefont {De~Nardis}, \citenamefont {Collura}, \citenamefont {Bertini},\
  and\ \citenamefont {Fagotti}}]{Piroli2017}%
  \BibitemOpen
  \bibfield  {author} {\bibinfo {author} {\bibfnamefont {L.}~\bibnamefont
  {Piroli}}, \bibinfo {author} {\bibfnamefont {J.}~\bibnamefont {De~Nardis}},
  \bibinfo {author} {\bibfnamefont {M.}~\bibnamefont {Collura}}, \bibinfo
  {author} {\bibfnamefont {B.}~\bibnamefont {Bertini}}, \ and\ \bibinfo
  {author} {\bibfnamefont {M.}~\bibnamefont {Fagotti}},\ }\href {\doibase
  10.1103/PhysRevB.96.115124} {\bibfield  {journal} {\bibinfo  {journal} {Phys.
  Rev. B}\ }\textbf {\bibinfo {volume} {96}},\ \bibinfo {pages} {115124}
  (\bibinfo {year} {2017})}\BibitemShut {NoStop}%
\bibitem [{\citenamefont {Takahashi}(1999)}]{TakTBA}%
  \BibitemOpen
  \bibfield  {author} {\bibinfo {author} {\bibfnamefont {M.}~\bibnamefont
  {Takahashi}},\ }\href {\doibase 10.1017/CBO9780511524332} {\emph {\bibinfo
  {title} {Thermodynamics of One-Dimensional Solvable Models}}}\ (\bibinfo
  {publisher} {Cambridge University Press},\ \bibinfo {year}
  {1999})\BibitemShut {NoStop}%
\bibitem [{\citenamefont {Bulchandani}\ \emph {et~al.}(2017)\citenamefont
  {Bulchandani}, \citenamefont {Vasseur}, \citenamefont {Karrasch},\ and\
  \citenamefont {Moore}}]{Solv}%
  \BibitemOpen
  \bibfield  {author} {\bibinfo {author} {\bibfnamefont {V.~B.}\ \bibnamefont
  {Bulchandani}}, \bibinfo {author} {\bibfnamefont {R.}~\bibnamefont
  {Vasseur}}, \bibinfo {author} {\bibfnamefont {C.}~\bibnamefont {Karrasch}}, \
  and\ \bibinfo {author} {\bibfnamefont {J.~E.}\ \bibnamefont {Moore}},\ }\href
  {\doibase 10.1103/PhysRevLett.119.220604} {\bibfield  {journal} {\bibinfo
  {journal} {Phys. Rev. Lett.}\ }\textbf {\bibinfo {volume} {119}},\ \bibinfo
  {pages} {220604} (\bibinfo {year} {2017})}\BibitemShut {NoStop}%
\bibitem [{\citenamefont {Gobert}\ \emph {et~al.}(2005)\citenamefont {Gobert},
  \citenamefont {Kollath}, \citenamefont {Schollw\"ock},\ and\ \citenamefont
  {Sch\"utz}}]{Gobert}%
  \BibitemOpen
  \bibfield  {author} {\bibinfo {author} {\bibfnamefont {D.}~\bibnamefont
  {Gobert}}, \bibinfo {author} {\bibfnamefont {C.}~\bibnamefont {Kollath}},
  \bibinfo {author} {\bibfnamefont {U.}~\bibnamefont {Schollw\"ock}}, \ and\
  \bibinfo {author} {\bibfnamefont {G.}~\bibnamefont {Sch\"utz}},\ }\href
  {\doibase 10.1103/PhysRevE.71.036102} {\bibfield  {journal} {\bibinfo
  {journal} {Phys. Rev. E}\ }\textbf {\bibinfo {volume} {71}},\ \bibinfo
  {pages} {036102} (\bibinfo {year} {2005})}\BibitemShut {NoStop}%
\bibitem [{\citenamefont {Misguich}\ \emph {et~al.}(2017)\citenamefont
  {Misguich}, \citenamefont {Mallick},\ and\ \citenamefont
  {Krapivsky}}]{Misguich}%
  \BibitemOpen
  \bibfield  {author} {\bibinfo {author} {\bibfnamefont {G.}~\bibnamefont
  {Misguich}}, \bibinfo {author} {\bibfnamefont {K.}~\bibnamefont {Mallick}}, \
  and\ \bibinfo {author} {\bibfnamefont {P.~L.}\ \bibnamefont {Krapivsky}},\
  }\href {\doibase 10.1103/PhysRevB.96.195151} {\bibfield  {journal} {\bibinfo
  {journal} {Phys. Rev. B}\ }\textbf {\bibinfo {volume} {96}},\ \bibinfo
  {pages} {195151} (\bibinfo {year} {2017})}\BibitemShut {NoStop}%
\bibitem [{\citenamefont {Collura}\ \emph {et~al.}(2018)\citenamefont
  {Collura}, \citenamefont {De~Luca},\ and\ \citenamefont {Viti}}]{DW}%
  \BibitemOpen
  \bibfield  {author} {\bibinfo {author} {\bibfnamefont {M.}~\bibnamefont
  {Collura}}, \bibinfo {author} {\bibfnamefont {A.}~\bibnamefont {De~Luca}}, \
  and\ \bibinfo {author} {\bibfnamefont {J.}~\bibnamefont {Viti}},\ }\href
  {\doibase 10.1103/PhysRevB.97.081111} {\bibfield  {journal} {\bibinfo
  {journal} {Phys. Rev. B}\ }\textbf {\bibinfo {volume} {97}},\ \bibinfo
  {pages} {081111} (\bibinfo {year} {2018})}\BibitemShut {NoStop}%
\bibitem [{\citenamefont {Bulchandani}\ and\ \citenamefont
  {Karrasch}(2019)}]{Subdiff}%
  \BibitemOpen
  \bibfield  {author} {\bibinfo {author} {\bibfnamefont {V.~B.}\ \bibnamefont
  {Bulchandani}}\ and\ \bibinfo {author} {\bibfnamefont {C.}~\bibnamefont
  {Karrasch}},\ }\href {\doibase 10.1103/PhysRevB.99.121410} {\bibfield
  {journal} {\bibinfo  {journal} {Phys. Rev. B}\ }\textbf {\bibinfo {volume}
  {99}},\ \bibinfo {pages} {121410} (\bibinfo {year} {2019})}\BibitemShut
  {NoStop}%
\bibitem [{\citenamefont {St{\'e}phan}(2019)}]{Stephan}%
  \BibitemOpen
  \bibfield  {author} {\bibinfo {author} {\bibfnamefont {J.-M.}\ \bibnamefont
  {St{\'e}phan}},\ }\href {\doibase 10.21468/SciPostPhys.6.5.057} {\bibfield
  {journal} {\bibinfo  {journal} {SciPost Phys.}\ }\textbf {\bibinfo {volume}
  {6}},\ \bibinfo {pages} {57} (\bibinfo {year} {2019})}\BibitemShut {NoStop}%
\bibitem [{\citenamefont {Gamayun}\ \emph {et~al.}(2019)\citenamefont
  {Gamayun}, \citenamefont {Miao},\ and\ \citenamefont {Ilievski}}]{LL1}%
  \BibitemOpen
  \bibfield  {author} {\bibinfo {author} {\bibfnamefont {O.}~\bibnamefont
  {Gamayun}}, \bibinfo {author} {\bibfnamefont {Y.}~\bibnamefont {Miao}}, \
  and\ \bibinfo {author} {\bibfnamefont {E.}~\bibnamefont {Ilievski}},\ }\href
  {\doibase 10.1103/PhysRevB.99.140301} {\bibfield  {journal} {\bibinfo
  {journal} {Phys. Rev. B}\ }\textbf {\bibinfo {volume} {99}},\ \bibinfo
  {pages} {140301} (\bibinfo {year} {2019})}\BibitemShut {NoStop}%
\bibitem [{\citenamefont {Misguich}\ \emph {et~al.}(2019)\citenamefont
  {Misguich}, \citenamefont {Pavloff},\ and\ \citenamefont {Pasquier}}]{LL2}%
  \BibitemOpen
  \bibfield  {author} {\bibinfo {author} {\bibfnamefont {G.}~\bibnamefont
  {Misguich}}, \bibinfo {author} {\bibfnamefont {N.}~\bibnamefont {Pavloff}}, \
  and\ \bibinfo {author} {\bibfnamefont {V.}~\bibnamefont {Pasquier}},\ }\href
  {\doibase 10.21468/SciPostPhys.7.2.025} {\bibfield  {journal} {\bibinfo
  {journal} {SciPost Phys.}\ }\textbf {\bibinfo {volume} {7}},\ \bibinfo
  {pages} {25} (\bibinfo {year} {2019})}\BibitemShut {NoStop}%
\bibitem [{\citenamefont {Faddeev}\ \emph {et~al.}(2007)\citenamefont
  {Faddeev}, \citenamefont {Reyman},\ and\ \citenamefont
  {Takhtajan}}]{Faddeev}%
  \BibitemOpen
  \bibfield  {author} {\bibinfo {author} {\bibfnamefont {L.}~\bibnamefont
  {Faddeev}}, \bibinfo {author} {\bibfnamefont {A.}~\bibnamefont {Reyman}}, \
  and\ \bibinfo {author} {\bibfnamefont {L.}~\bibnamefont {Takhtajan}},\ }\href
  {https://books.google.com/books?id=0f-RCEwQhR8C} {\emph {\bibinfo {title}
  {Hamiltonian Methods in the Theory of Solitons}}},\ Classics in Mathematics\
  (\bibinfo  {publisher} {Springer Berlin Heidelberg},\ \bibinfo {year}
  {2007})\BibitemShut {NoStop}%
\bibitem [{\citenamefont {Arzamasovs}\ \emph {et~al.}(2014)\citenamefont
  {Arzamasovs}, \citenamefont {Bovo},\ and\ \citenamefont
  {Gangardt}}]{Gangardt}%
  \BibitemOpen
  \bibfield  {author} {\bibinfo {author} {\bibfnamefont {M.}~\bibnamefont
  {Arzamasovs}}, \bibinfo {author} {\bibfnamefont {F.}~\bibnamefont {Bovo}}, \
  and\ \bibinfo {author} {\bibfnamefont {D.~M.}\ \bibnamefont {Gangardt}},\
  }\href {\doibase 10.1103/PhysRevLett.112.170602} {\bibfield  {journal}
  {\bibinfo  {journal} {Phys. Rev. Lett.}\ }\textbf {\bibinfo {volume} {112}},\
  \bibinfo {pages} {170602} (\bibinfo {year} {2014})}\BibitemShut {NoStop}%
\bibitem [{\citenamefont {Vasseur}\ and\ \citenamefont
  {Moore}(2016)}]{VasseurMoore}%
  \BibitemOpen
  \bibfield  {author} {\bibinfo {author} {\bibfnamefont {R.}~\bibnamefont
  {Vasseur}}\ and\ \bibinfo {author} {\bibfnamefont {J.~E.}\ \bibnamefont
  {Moore}},\ }\href {\doibase 10.1088/1742-5468/2016/06/064010} {\bibfield
  {journal} {\bibinfo  {journal} {J. Stat. Mech.: Theory Exp}\ }\textbf
  {\bibinfo {volume} {2016}},\ \bibinfo {pages} {064010} (\bibinfo {year}
  {2016})}\BibitemShut {NoStop}%
\bibitem [{\citenamefont {Kulkarni}\ \emph {et~al.}(2015)\citenamefont
  {Kulkarni}, \citenamefont {Huse},\ and\ \citenamefont {Spohn}}]{Kulk2}%
  \BibitemOpen
  \bibfield  {author} {\bibinfo {author} {\bibfnamefont {M.}~\bibnamefont
  {Kulkarni}}, \bibinfo {author} {\bibfnamefont {D.~A.}\ \bibnamefont {Huse}},
  \ and\ \bibinfo {author} {\bibfnamefont {H.}~\bibnamefont {Spohn}},\ }\href
  {\doibase 10.1103/PhysRevA.92.043612} {\bibfield  {journal} {\bibinfo
  {journal} {Phys. Rev. A}\ }\textbf {\bibinfo {volume} {92}},\ \bibinfo
  {pages} {043612} (\bibinfo {year} {2015})}\BibitemShut {NoStop}%
\bibitem [{\citenamefont {Lakshmanan}\ \emph {et~al.}(1976)\citenamefont
  {Lakshmanan}, \citenamefont {Ruijgrok},\ and\ \citenamefont
  {Thompson}}]{Lakshmanan}%
  \BibitemOpen
  \bibfield  {author} {\bibinfo {author} {\bibfnamefont {M.}~\bibnamefont
  {Lakshmanan}}, \bibinfo {author} {\bibfnamefont {T.}~\bibnamefont
  {Ruijgrok}}, \ and\ \bibinfo {author} {\bibfnamefont {C.}~\bibnamefont
  {Thompson}},\ }\href {\doibase https://doi.org/10.1016/0378-4371(76)90106-0}
  {\bibfield  {journal} {\bibinfo  {journal} {Physica A}\ }\textbf {\bibinfo
  {volume} {84}},\ \bibinfo {pages} {577 } (\bibinfo {year}
  {1976})}\BibitemShut {NoStop}%
\bibitem [{Note2()}]{Note2}%
  \BibitemOpen
  \bibinfo {note} {Stationary states with characteristic length scale
  $l_{\protect \mathbf {\Omega }}$ and average spin $\mu /2$ along a given axis
  are helices with $\tau =\mu /l_\protect \mathbf {\Omega }$ and $\protect
  \mathcal {E} = (1-\mu ^2)/2l_{\protect \mathbf {\Omega }}^2$.}\BibitemShut
  {Stop}%
\bibitem [{\citenamefont {Das}\ \emph {et~al.}(2018{\natexlab{a}})\citenamefont
  {Das}, \citenamefont {Chakrabarty}, \citenamefont {Dhar}, \citenamefont
  {Kundu}, \citenamefont {Huse}, \citenamefont {Moessner}, \citenamefont
  {Ray},\ and\ \citenamefont {Bhattacharjee}}]{ClassXXX}%
  \BibitemOpen
  \bibfield  {author} {\bibinfo {author} {\bibfnamefont {A.}~\bibnamefont
  {Das}}, \bibinfo {author} {\bibfnamefont {S.}~\bibnamefont {Chakrabarty}},
  \bibinfo {author} {\bibfnamefont {A.}~\bibnamefont {Dhar}}, \bibinfo {author}
  {\bibfnamefont {A.}~\bibnamefont {Kundu}}, \bibinfo {author} {\bibfnamefont
  {D.~A.}\ \bibnamefont {Huse}}, \bibinfo {author} {\bibfnamefont
  {R.}~\bibnamefont {Moessner}}, \bibinfo {author} {\bibfnamefont {S.~S.}\
  \bibnamefont {Ray}}, \ and\ \bibinfo {author} {\bibfnamefont
  {S.}~\bibnamefont {Bhattacharjee}},\ }\href {\doibase
  10.1103/PhysRevLett.121.024101} {\bibfield  {journal} {\bibinfo  {journal}
  {Phys. Rev. Lett.}\ }\textbf {\bibinfo {volume} {121}},\ \bibinfo {pages}
  {024101} (\bibinfo {year} {2018}{\natexlab{a}})}\BibitemShut {NoStop}%
\bibitem [{\citenamefont {Das}\ \emph {et~al.}(2018{\natexlab{b}})\citenamefont
  {Das}, \citenamefont {Damle}, \citenamefont {Dhar}, \citenamefont {Huse},
  \citenamefont {Kulkarni}, \citenamefont {Mendl},\ and\ \citenamefont
  {Spohn}}]{ClassXXZ}%
  \BibitemOpen
  \bibfield  {author} {\bibinfo {author} {\bibfnamefont {A.}~\bibnamefont
  {Das}}, \bibinfo {author} {\bibfnamefont {K.}~\bibnamefont {Damle}}, \bibinfo
  {author} {\bibfnamefont {A.}~\bibnamefont {Dhar}}, \bibinfo {author}
  {\bibfnamefont {D.~A.}\ \bibnamefont {Huse}}, \bibinfo {author}
  {\bibfnamefont {M.}~\bibnamefont {Kulkarni}}, \bibinfo {author}
  {\bibfnamefont {C.~B.}\ \bibnamefont {Mendl}}, \ and\ \bibinfo {author}
  {\bibfnamefont {H.}~\bibnamefont {Spohn}},\ }\href@noop {} {\  (\bibinfo
  {year} {2018}{\natexlab{b}})},\ \Eprint {http://arxiv.org/abs/1901.00024}
  {arXiv:1901.00024 [cond-mat.stat-mech]} \BibitemShut {NoStop}%
\bibitem [{\citenamefont {Borsi}\ \emph {et~al.}(2019)\citenamefont {Borsi},
  \citenamefont {Pozsgay},\ and\ \citenamefont {Pristyák}}]{PG1}%
  \BibitemOpen
  \bibfield  {author} {\bibinfo {author} {\bibfnamefont {M.}~\bibnamefont
  {Borsi}}, \bibinfo {author} {\bibfnamefont {B.}~\bibnamefont {Pozsgay}}, \
  and\ \bibinfo {author} {\bibfnamefont {L.}~\bibnamefont {Pristyák}},\
  }\href@noop {} {\  (\bibinfo {year} {2019})},\ \Eprint
  {http://arxiv.org/abs/1908.07320} {arXiv:1908.07320 [cond-mat.stat-mech]}
  \BibitemShut {NoStop}%
\bibitem [{\citenamefont {Pozsgay}(2019)}]{PG2}%
  \BibitemOpen
  \bibfield  {author} {\bibinfo {author} {\bibfnamefont {B.}~\bibnamefont
  {Pozsgay}},\ }\href@noop {} {\  (\bibinfo {year} {2019})},\ \Eprint
  {http://arxiv.org/abs/1910.12833} {arXiv:1910.12833 [cond-mat.stat-mech]}
  \BibitemShut {NoStop}%
\bibitem [{\citenamefont {Medenjak}\ \emph {et~al.}(2019)\citenamefont
  {Medenjak}, \citenamefont {Nardis},\ and\ \citenamefont {Yoshimura}}]{Subs}%
  \BibitemOpen
  \bibfield  {author} {\bibinfo {author} {\bibfnamefont {M.}~\bibnamefont
  {Medenjak}}, \bibinfo {author} {\bibfnamefont {J.~D.}\ \bibnamefont
  {Nardis}}, \ and\ \bibinfo {author} {\bibfnamefont {T.}~\bibnamefont
  {Yoshimura}},\ }\href@noop {} {\  (\bibinfo {year} {2019})},\ \Eprint
  {http://arxiv.org/abs/1911.01995} {arXiv:1911.01995 [cond-mat.stat-mech]}
  \BibitemShut {NoStop}%
\end{thebibliography}%
\end{document}